\documentclass[aps,twocolumn,prl,groupedaddress,showpacs]{revtex4-1}

\usepackage{graphicx,float}
\usepackage[normalem]{ulem}
\usepackage{color}

\newcommand{\Niso}{$^{14}$N }

\usepackage{calligra}
\DeclareMathAlphabet{\mathcalligra}{T1}{calligra}{m}{n}
\DeclareFontShape{T1}{calligra}{m}{n}{<->s*[2.2]callig15}{}

\def\be{\begin{equation}}
\def\ee{\end{equation}}
\def\e#1{\label{#1}\end{equation}}
\def\bea{\begin{eqnarray}}
\def\eea{\end{eqnarray}}
\def\ea#1{\label{#1}\end{eqnarray}}

\def\bem#1{\begin{mathletters}\label{#1}}
\def\eml{\end{mathletters}}

\def\ket#1{{|#1\rangle}}

\def\4#1{{\boldsymbol{#1}}}
\def\8#1{{\widetilde{#1}}}
\def\bse{\begin{subequations}}
\def\ese{\end{subequations}}
\def\nn{\nonumber}

\def\Rb87{$^{87}\text{Rb}$}
\def\0{\ket{0}}
\def\1{\ket{1}}

\begin{document}
\title{Generation of entangled photon strings using NV centers in diamond}
\author{D. D. Bhaktavatsala Rao }
\author{Sen Yang}
\author{J\"{o}rg Wrachtrup}

\affiliation{3. Physikalisches Institut, Research Center SCOPE, and MPI for Solid State Research, University of Stuttgart, Pfaffenwaldring 57, 70569 Stuttgart, Germany
}%
\date{\today}
\begin{abstract}
We present a scheme to generate entangled photons using the NV centers in diamond. We show how the long-lived nuclear spin in diamond can mediate entanglement between multiple photons thereby increasing the length of entangled photon string.
With the proposed scheme one could generate both n-photon GHZ and cluster states.  We present an experimental scheme realizing the same and estimating the rate of entanglement generation both in the presence and absence of a cavity.
 \end{abstract}
\maketitle
With controlled generation and manipulation, quantum states of light are efficient carriers of quantum information with applications in quantum computing (QC), communications and cryptography. One of the major drawbacks with optical quantum information processing (QIP) is the absence of suitable nonlinear interactions to realize universal quantum gates, for example a CNOT gate between two photonic qubits. To overcome this difficulty one may choose the implementation of desired quantum computation through a one-way quantum computer model \cite{oneway} which requires the initialization of the quantum register in a globally entangled cluster state. The computation is then followed by performing only single qubit measurements. The one-way quantum computer or the measurement-based quantum computation using photonic qubits (polarization states) has already been shown to be a fault-tolerant model for QC and is tolerant to qubit losses \cite{losstol}. The main hurdle in realizing optical QIP using this scheme is the generation of multi-qubit cluster state, the key initialization step of the model. While the experimental implementations succeeded to generate 6-qubit photonic cluster state optically \cite{cluster6}, scaling this number further is not so clear. To this end there have been proposals to use solid-state emitters such as a periodically pumped quantum dot (QD) for the generation of a one-dimensional cluster states \cite{qd1,qd2}. In this work we consider another possible solid-state system, the NV centers in diamond, to generate multi-photon entangled states.

The NV center provides a hybrid spin system in which electron spins are used for fast \cite{Awschalom_fast_control}, high-fidelity control \cite{Wrachtrup_opt_control} and readout \cite{Neumann_singleshot,Hanson_singleshot}, and nuclear spins are well-isolated from their environment yielding ultra-long coherence time \cite{Lukin_C13}. Electron and nuclear spins could form a small-scale quantum register \cite{Wrachtrup_error_correction,Wrachtrup_pairs,Hanson_error_correction} allowing for e.g. necessary high-fidelity quantum error correction \cite{Wrachtrup_error_correction}. Furthermore, the NV electron spin can be entangled with an emitted optical photon \cite{Lukin_photon,kosaka_entangled} and further quantum entanglement \cite{Hanson_entanglement} and quantum teleportation \cite{Hanson_teleportation} between two remote NV centers have already been experimentally demonstrated. We have also recently demonstrated the ability of this solid-state device to store quantum information from a light field into the defect spins and a repetitive readout of the memory, essential for scalable networks. In addition there have been other proposals to create large scalable QIP in diamond using a photonic architecture \cite{nuclearcluster} where cluster/topological states of the long-lived nuclear spins in various defect centers are generated using photons. Here we show the reverse where the nuclear spin of a single defect center can mediate the entanglement between photons thereby generating large strings of entangled photons.
\begin{figure*}
\begin{center}
\includegraphics[width=0.68\textwidth]{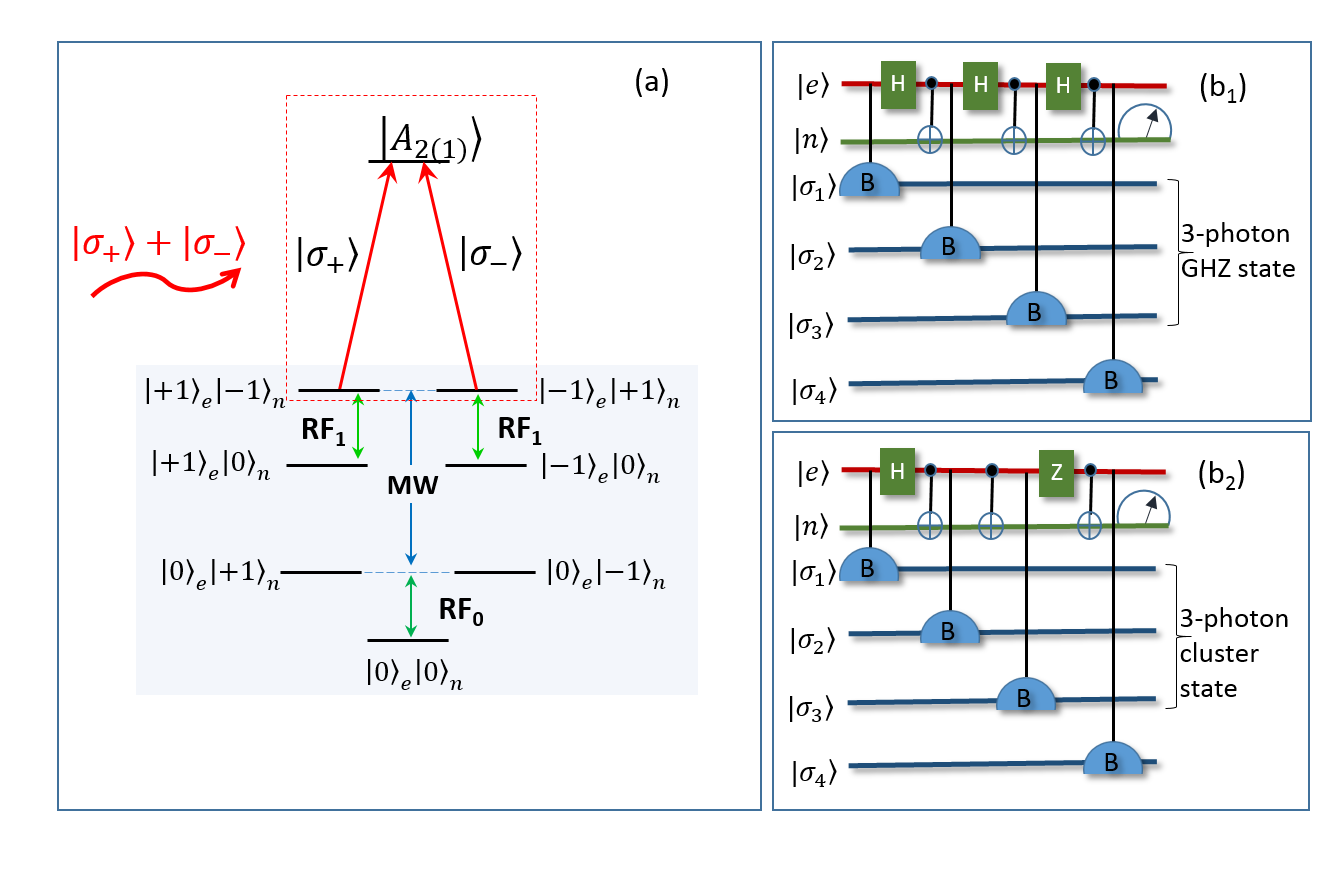}
\end{center}
\label{level}
\vspace{-5mm}
\caption{  \textbf{a,}  The relevant level structure of the NV, with excited state $\ket{A_{2(1)}}=\ket{E_-}\ket{+1}_e \pm \ket{E_+}\ket{-1}_e$, and ground states $\ket{E_0}\otimes\ket{m_s}_e\ket{m_n}_n$, where $\ket{E_{0,\pm}}$ are orbital states with angular momentum projection $0,\pm1$ respectively. $\ket{m_s=0,\pm1}_e$ and $\ket{m_n=0,\pm1}_n$ are corresponding eigenstates of the electron spin and the \Niso nuclear spin. The spatial part of the wave function $\ket{E_{0}}$ is not explicitly written for simplicity when referring to ground states. The $\Lambda$ system is highlighted. The individual manipulation of electron and nuclear spins by the microwave (MW) and radio-frequency (RF) fields is also shown in the figure.
    \textbf{b}, Circuit diagram for generating a 3-photon entangled state via the nuclear spin. $\text{H}$ indicates the Hadamard gate on the electron spin in the two-level basis $\ket{\pm}_e$, $X$ and $Y$ are the standard controlled Pauli operations performed on the nuclear spin conditioned on the state of the electron. The interaction of the electron spin with the incoming photon is represented as a Bell type measurement ($B$) on the electron-photon system projecting it onto the entangled state $\frac{1}{\sqrt{2}}[\ket{+1}_e\ket{\sigma_-} + \ket{-1}_e\ket{\sigma_+}] $. Finally a measurement on the nuclear spin will project the $N$-photon state onto to the GHZ ($b_1$) or cluster state ($b_2$) as detailed in the text.}
\end{figure*}

The basic element of our system is a single NV center consisting of an electronic spin (S=1) and intrinsic \Niso nuclear spin (I=1), coupled by hyperfine interaction. The interaction with optical photons in a $\Lambda$ system forms the basis of our scheme, and is shown Fig.~\ref{level} (a). The three-level $\Lambda$ system is formed by the two ground states of the electron $\{\ket{+1}_e,\ket{-1}_e$ and an excited state $\ket{A_2(1)}$. Owing to zero magnetic moment of the electron spin in the $\ket{A_2}$ state~\cite{Maze_NV_structure} and total angular momentum conservation, both ground states can be excited to the same state $\ket{A_2(1)}$ through absorption of a photon with $\sigma_{+}$ and $\sigma_{-}$ polarization respectively. We start with the two ground states being degenerate and the NV spin system prepared in superposition state $\ket{\Psi^{+}}=\frac{1}{\sqrt{2}}(\ket{+1}_e + \ket{-1}_e)$ . A photon in state $\frac{1}{\sqrt{2}}(\ket{\sigma_+}+ \ket{\sigma_-})$ and in resonance with the $A_2$ transition is sent into the NV center. After absorption of a photon, the collective photon-NV spin system is projected into the state $\ket{A_2}$, and hence the electron photon state after subsequent emission would remain in the entangled state 
$\ket{\psi_e^{(1)}}= \frac{1}{\sqrt{2}}[\ket{+1}_e\ket{\sigma_-} + \ket{-1}_e\ket{\sigma_+}] $. 
Since the excitation process is like a projective measurement in the excited state basis, re-exciting the NV-system by a second photon pulse would immediately disentangle the first photon and the total state after the emission would $\ket{\psi_e^{(2)}}= \frac{1}{\sqrt{2}}[\ket{+1}_e\ket{\sigma_-} + \ket{-1}\ket{\sigma_+}]\otimes \frac{1}{\sqrt{2}}(\ket{\sigma_+}+\ket{\sigma_-})$. Hence by post selecting only the absorption events one can see that the electron spin alone cannot mediate the entanglement between multiple photons as found in other solid-state proposals \cite{qd1}. For this we use the hyperfine interaction with its nuclear spin to transfer the entanglement of the electron with the emitted photons to the nuclear spin as detailed below.

After the absorption and emission of first photon the electron spin gets entangled to the photon, $\ket{\psi_e^{(1)}}$ as described above. Before re-pumping the NV system with the second photon we perform a C-NOT between the NV electron spin (see Fig. 1(b)) and its intrinsic ${}^{14}N$ spin, thus entangling them, viz.,  $\ket{\psi_{en}^{(1)}}= \frac{1}{\sqrt{2}}[\ket{+1}_e\ket{\sigma_-}\ket{+1}_n \pm \ket{-1}_e\ket{\sigma_+}\ket{-1}_n]$, where $\ket{\pm 1}_n$ are the basis states of the nuclear spin. A subsequent absorption and emission of the photon would leave the total system in the state, $\ket{\psi_{en}^{(2)}} = \ket{\psi_e^{(2)}}\ket{\psi_n^{(1)}}$, where the electron is now entangled with the second photon and the nuclear spin entangled to the first photon. Following the circuit diagram in Fig. 1($b_1$), and after the absorption of the third photon the nuclear spin and the two photons are now projected on to the tripartite entangled state where the two photons are in the maximally entangled GHZ state for any spin projection of the nuclear spin viz..,
\be
\ket{\phi_{n}^{(2)}}  = \frac{1}{\sqrt{2}}[\ket{+1}_n\ket{G^{(2)}_-} + \ket{-1}_n\ket{G^{(2)}_+}].
\ee
where $\ket{{G^{(2)}_\mp}} = \frac{1}{\sqrt{2}}[\ket{\sigma_+}\ket{\sigma_+} \mp \ket{\sigma_-}\ket{\sigma_-}]$.
Continuing this procedure further as shown in Fig. 1($b_1$), a $n$-photon GHZ state is generated after the electron has been excited by a $(n+1)^{th}$  photon. 

Instead if we manipulate the electron spin with different local operations as shown in Fig. 1($b_2$) we can for example project the nuclear spin and the two photons onto a different tripartite entangled state as shown below after the absorption of the third photon. 
\be
\ket{\phi_{n}^{(2)}}  = \frac{1}{\sqrt{2}}[\ket{+1}_n\ket{C^{(2)}_0} + \ket{-1}_n\ket{C^{(2)}_1}].
\ee
Rewriting $\ket{\sigma_{\mp}}$ as $\ket{0(1)}$, $C^{(2)}_{0(1)}$ one can see that we have created a two-photon cluster state \cite{cluster}
\bea
\ket{C^{(2)}_0} &=& \frac{1}{2}\bigotimes_{a=1}^2(\ket{0}_aZ^{(a+1)} + \ket{1}_a) \nn \\
\ket{C^{(2)}_1} &=& \frac{1}{2}\bigotimes_{a=1}^2(\ket{0}_a + \ket{1}_aZ^{(a+1)}),
\eea
where $Z$ is the Pauli operator. Here we would like to highlight the possibility that by controlling the solid-state spins we could project the $n$-photon state onto different entangled states.

The proposed scheme can be implemented efficiently at low temperatures (T$<$8 K) in a low strain ($\approx$1.2 GHz) NV center aligned along [111] direction\cite{Lukin_photon}.  At such temperatures the optical transitions are well resolved allowing for resonant excitation to perform efficient initialization and projective high-fidelity single shot readout on the electron spin. The degeneracy of the ground states can be maintained by switching off any external magnetic field. In addition we can maintain the coherence of the nuclear spin for about a minute a key parameter in our proposal. In solid-state proposals using electrons spin the rapid decoherence of the electronic spin could be a serious bottleneck for scalable production of entangled photons, which we could overcome in our present protocol. Due to the finite operation time for entangling each outgoing photon with the nuclear spin there is a time delay $\tau$ between any two subsequent emissions. This immediately requires the emitted photon to be stored in a single mode optical fiber at the mode frequency of the Zero Phonon Line (ZPL) of NV.  Due to long $T_1$ the nuclear spin does not flip and hence the emitted photon remains entangled with it till the next photon arrives. 

The next important factor to consider is the case where the electron spin is in the dark state to the applied laser field. To see this one can rewrite the entangled state of the total system ,$ \frac{1}{\sqrt{2}}[\ket{+1}_e\ket{\phi^{(+1)}_{n,\sigma}}+ \ket{-1}_e\ket{\tilde{\phi}^{(-1)}_{n,\sigma} }]$, in the electron's bright and dark state basis ($\ket{b(d)}_e = \frac{1}{\sqrt{2}}[\ket{+1}_e\pm\ket{-1}_e]$) ,  as $ \frac{1}{\sqrt{2}}[\ket{b}_e\ket{\psi^{(+1)}_{n,\sigma}}+  \ket{d}_e\ket{\psi^{(-1)}_{n,\sigma} }]$, where ${\psi^{(\pm 1)}_{n,\sigma}}$ are the bright and dark combinations of ${\phi^{(+1)}_{n,\sigma}}$. Clearly there is a $50\%$chance of not having a resonant absorption at any time $t$. This makes the situation probabilistic as for $N$-photons incident on the $NV$ the probability with which any $m$-photons are resonantly absorbed has a Gaussian distribution $P(n) = \frac{N!}{(N-n)!(N+n)!} \approx \frac{2}{\sqrt{\pi N}}{\rm e}^{-\frac{2(N-n)^2}{N}}$ centered around $N/2$ with width $\sqrt{N/2}$. Also the $m$-photon state obtained for any two repetitive cycles would be different, reducing the average fidelity of the $m$-photon entangled state. 

To overcome this problem and only obtain a $m$-photon state consistently by only bright state absorptions we switch on two lasers allowing for an individual excitation on the transitions first to $\ket{\pm 1} \leftrightarrow \ket{A_1}$ followed by the entangling transition $\ket{\pm 1} \leftrightarrow \ket{A_2}$. If the electron is in the wrong state  (dark state) then the $\ket{A_1}$ transition will populate it to the $\ket{0}_e$ state, thereby ending the operation cycle as no more absorption is possible (by this we eliminate the errors due to the dark state evolution in the presence of gate operations and hyperfine coupling that could mix it with the bright state). Instead if the electron is in the correct state (bright state) it does not get excited to $\ket{A_1}$ and absorbs a photon resonant with $\ket{A_2}$ transition. The probability of having $m$ photons entangled at the end of an operation cycle is given by
\be
P_0(m) =\exp(-(m-1)\ln 2),
\ee
for $N \gg m$. For example one can choose the time for an operation cycle $t = N\tau$ ($\tau$ is the separation between any two subsequent photons incident on the NV). With a typical time scale of $\tau = 1\mu$s, one may choose an operation time of $t=100\mu$s, so that in $1000$ repetitions there will be $\sim 62$ events where the minimum length of the entangled chain of photons would be $5$, and two events with a minimum length of the entangled chain of photons being $10$. These numbers are estimated for $100\%$ collection efficiencies. For example in current experimental setup without a cavity, the probability to observe an entangled state with at least two-photons is $\sim 100$s. Such low efficiency is also reported in the heralded entanglement with two NV's, where an entanglement event has been detected for every $2$ minutes \cite{Hanson_entanglement}.  

To get closer to the ideal estimate for the number of entangled photons one can use a cavity to boost the emission into ZPL. For example it was recently shown in NV coupled photonic crystal cavity experiment \cite{cavity} that in the presence of a cavity $70\%$ of the emission would be in the ZPL, and with an achievable collection efficiency of $\sim 90\%$ ZPL photons we achieve a $10$-photon entangled state per second. 

We now estimate the fidelity of the $m$-photon entangled state generated in the presence of various errors arising from imperfect gates operations and by dephasing of the electron/nuclear spin coherence by the surrounding spin-bath. Though the nuclear spin coherence time is limited by the $T_1$ time of the electron, one should also consider the decoherence effects arising from the direct nuclear coupling with the surrounding spins as the nuclear spin coherence should survive for few minutes if one has to achieve entanglement larger number of photons.  We now analyze these errors individually and compare them. 

During the protocol we perform multiple Hadamard and C-NOT gate operations (see Fig. 1(b)) which entangles (mixes) the basis states of electron and nuclear spins. Any random phase obtained by these spins are eventually transferred into the $m$-photon state as the photons are always entangled to the nuclear spin. To see this we shall consider the case where the nuclear spin obtains a random-phase $\eta$ during the $\ket{A_2(1)}$ transition. Due to this the two-photon entangled state given in Eq. (1) gets modified to
\bea
\ket{\phi_{n}^{(2)}}_{err}  &=& \frac{1}{\sqrt{2}}[\ket{+1}_n(\cos\eta \ket{G^{(2)}_-} +\sin\eta \ket{\tilde{G}^{(2)}_-}) \nn \\ &+& \ket{-1}_n(\cos\eta \ket{G^{(2)}_+} +\sin\eta \ket{\tilde{G}^{(2)}_-})].
\eea
where $\ket{{G^{(2)}_\mp}} = \frac{1}{\sqrt{2}}[\ket{\sigma_+}\ket{\sigma_-} \mp \ket{\sigma_-}\ket{\sigma_+}]$.

Imperfect gate operations would lead to both phase and amplitude errors and can a dominant role in the fidelity loss.
One can see this from Fig.2 we have plotted the fidelity ${F}_m = Tr[\rho^{ideal}_{m\sigma}\rho_{m\sigma}]$ as a function of $m$ (entangled photons) in the presence of errors.  For gate operations errors are introduced by imperfect unitary rotations that realize the Hadamard and CNOT gates.  We would also like to note that for the present physical system imperfect Hadamard gate will also lead to population loss outside the qubit subspace as it is performed indirectly via the $\ket{0}_e$ state \cite{holonomic}. This will not directly effect the fidelity but decreases the event rate as discussed earlier.


\begin{figure}
\hspace{-10mm}
\includegraphics[width=100mm]{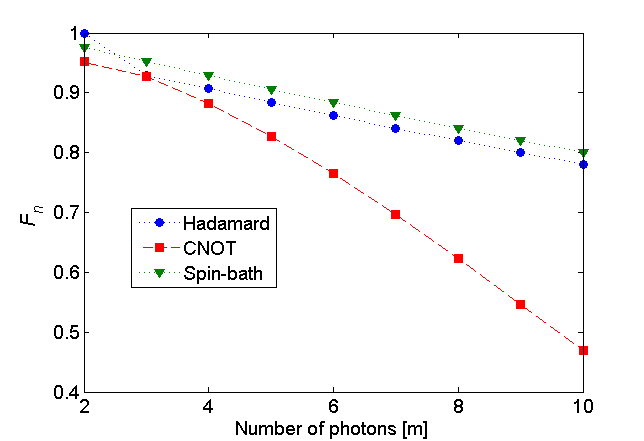}
\label{fds}
\caption{The fidelity of ${F}_m = Tr[\rho^{ideal}_{m\sigma}\rho_{m\sigma}]$ of the $n$-photon GHZ state generated by the solid-state spins is plotted as a function of $m$ both in the presence of errors while performing the Hadamard, C-NOT gates and due to a random phase obtained by the nuclear spin at each interval $\tau$ due to the quasi-static spin bath. Both the rotation angle errors for the unitary gates and the phase error due to spin-bath is chosen to take a maximum value of $10^{o}$. } 
\end{figure}

In addition to the unitary pulse errors, there are other non-unitary errors which would arise as the electron and nuclear spin are coupled to a spin-bath comprised of the surrounding ${}^{13}C$ nuclear spins. The \Niso nuclear spin is continuously evolving (dominant phase evolution) under the hyperfine (dipolar) coupling with both the spin-$1$ (electron) and spin-$1/2$ (${}^{13}C$ nuclear spins). 
Owing to the large quadrupolar splitting of the nuclear spin and the large zero field splitting of the electron spin, all the non-secular terms that induce flips of the electron and nitrogen nuclear spin can be safely neglected, and the Hamiltonian of the whole system in the interaction picture can be simplified to a pure dephasing model
\begin{equation}
\label{Hamil}
H = A I_{z}S_{z} + \frac{\gamma_n}{\gamma_e}I_{z}\sum_{j} {\textbf{\textit{A}}_{e,j,z}\cdot\textbf{I}^{j}}
+ S_{z}\sum_{j} {\textbf{\textit{A}}_{e,j,z}\cdot\textbf{I}^{j}},
\end{equation}
where the hyperfine coupling tensor of describing the interaction with the $j^{th}$-bath spin located at ${r}_{ej}$ is given by $\textbf{\textit{A}}_{e,j} = \frac{\mu_0}{4\pi}\frac{\gamma_e\gamma_c}{r_{ej}^3}(1-3\frac{\textbf{r}_{ej}\textbf{r}_{ej}}{r_{ej}^2})$, and $A$ is the hyperfine coupling between the electron and \Niso nuclear spin.
Since the electron itself is interacting strongly with the ${}^{13}C$ spin-bath, the nuclear spin sees an effective bath interaction (gradient magnetic field)  through its hyperfine interaction with the electron spin along with the direct dipole-dipole interaction to the ${}^{13}C$ nuclear spin-bath (which is generally assumed to be weak). The additional phase gained by both the electron and nuclear spins due to their mutual hyperfine coupling is a more systematic error and can be canceled by a simple Hahn echo on the electron spin. Due to the quasi-static nature of the bath, the random phase obtained by the nuclear spin would reduce the entanglement fidelity as the phase enters into the photonic state after every absorption event. In Fig. 2 we show the reduction of the fidelity with increasing $m$.

In conclusion we show that the long-lived nuclear spins in diamond could be a resource to mediate (generate) entanglement between single photons. The long $T_{1(2)}$ times of these spins allow for a controlled creation of multi-photon GHZ or cluster states, which could be useful for QIP with photons. Due to a very small coupling to the surrounding spin-bath random phase errors have less harmful effects that the errors in gate operations of the same order.  We predicted the generation of an entangled photon string with a minimum length of $10$ photons per second in the presence of a cavity, where the emission of outgoing photons into the ZPL and the detection efficiencies are quite high. These solid-state spins both with their ability to store quantum information from the incoming photons and generate entangled pairs could have a promising impact to the field of quantum communications where on demand generation of high fidelity entangled photons and their storage is quintessential.

\bibliography{references}

\begin{thebibliography}{22}%
\makeatletter
\providecommand \@ifxundefined [1]{%
 \@ifx{#1\undefined}
}%
\providecommand \@ifnum [1]{%
 \ifnum #1\expandafter \@firstoftwo
 \else \expandafter \@secondoftwo
 \fi
}%
\providecommand \@ifx [1]{%
 \ifx #1\expandafter \@firstoftwo
 \else \expandafter \@secondoftwo
 \fi
}%
\providecommand \natexlab [1]{#1}%
\providecommand \enquote  [1]{``#1''}%
\providecommand \bibnamefont  [1]{#1}%
\providecommand \bibfnamefont [1]{#1}%
\providecommand \citenamefont [1]{#1}%
\providecommand \href@noop [0]{\@secondoftwo}%
\providecommand \href [0]{\begingroup \@sanitize@url \@href}%
\providecommand \@href[1]{\@@startlink{#1}\@@href}%
\providecommand \@@href[1]{\endgroup#1\@@endlink}%
\providecommand \@sanitize@url [0]{\catcode `\\12\catcode `\$12\catcode
  `\&12\catcode `\#12\catcode `\^12\catcode `\_12\catcode `\%12\relax}%
\providecommand \@@startlink[1]{}%
\providecommand \@@endlink[0]{}%
\providecommand \url  [0]{\begingroup\@sanitize@url \@url }%
\providecommand \@url [1]{\endgroup\@href {#1}{\urlprefix }}%
\providecommand \urlprefix  [0]{URL }%
\providecommand \Eprint [0]{\href }%
\providecommand \doibase [0]{http://dx.doi.org/}%
\providecommand \selectlanguage [0]{\@gobble}%
\providecommand \bibinfo  [0]{\@secondoftwo}%
\providecommand \bibfield  [0]{\@secondoftwo}%
\providecommand \translation [1]{[#1]}%
\providecommand \BibitemOpen [0]{}%
\providecommand \bibitemStop [0]{}%
\providecommand \bibitemNoStop [0]{.\EOS\space}%
\providecommand \EOS [0]{\spacefactor3000\relax}%
\providecommand \BibitemShut  [1]{\csname bibitem#1\endcsname}%
\let\auto@bib@innerbib\@empty
\bibitem [{\citenamefont {Raussendorf}\ and\ \citenamefont
  {Briegel}(2001)}]{oneway}%
  \BibitemOpen
  \bibfield  {author} {\bibinfo {author} {\bibfnamefont {R.}~\bibnamefont
  {Raussendorf}}\ and\ \bibinfo {author} {\bibfnamefont {H.~J.}\ \bibnamefont
  {Briegel}},\ }\href {\doibase 10.1103/PhysRevLett.86.5188} {\bibfield
  {journal} {\bibinfo  {journal} {Phys. Rev. Lett.}\ }\textbf {\bibinfo
  {volume} {86}},\ \bibinfo {pages} {5188} (\bibinfo {year}
  {2001})}\BibitemShut {NoStop}%
\bibitem [{\citenamefont {Varnava}\ \emph {et~al.}(2006)\citenamefont
  {Varnava}, \citenamefont {Browne},\ and\ \citenamefont {Rudolph}}]{losstol}%
  \BibitemOpen
  \bibfield  {author} {\bibinfo {author} {\bibfnamefont {M.}~\bibnamefont
  {Varnava}}, \bibinfo {author} {\bibfnamefont {D.~E.}\ \bibnamefont {Browne}},
  \ and\ \bibinfo {author} {\bibfnamefont {T.}~\bibnamefont {Rudolph}},\ }\href
  {\doibase 10.1103/PhysRevLett.97.120501} {\bibfield  {journal} {\bibinfo
  {journal} {Phys. Rev. Lett.}\ }\textbf {\bibinfo {volume} {97}},\ \bibinfo
  {pages} {120501} (\bibinfo {year} {2006})}\BibitemShut {NoStop}%
\bibitem [{\citenamefont {Lu}(2007)}]{cluster6}%
  \BibitemOpen
  \bibfield  {author} {\bibinfo {author} {\bibfnamefont {C.-Y.}\ \bibnamefont
  {Lu}},\ }\href@noop {} {\bibfield  {journal} {\bibinfo  {journal} {{\it et
  al}, Nat Phys}\ }\textbf {\bibinfo {volume} {3}},\ \bibinfo {pages} {91}
  (\bibinfo {year} {2007})}\BibitemShut {NoStop}%
\bibitem [{\citenamefont {Lindner}\ and\ \citenamefont {Rudolph}(2009)}]{qd1}%
  \BibitemOpen
  \bibfield  {author} {\bibinfo {author} {\bibfnamefont {N.~H.}\ \bibnamefont
  {Lindner}}\ and\ \bibinfo {author} {\bibfnamefont {T.}~\bibnamefont
  {Rudolph}},\ }\href {\doibase 10.1103/PhysRevLett.103.113602} {\bibfield
  {journal} {\bibinfo  {journal} {Phys. Rev. Lett.}\ }\textbf {\bibinfo
  {volume} {103}},\ \bibinfo {pages} {113602} (\bibinfo {year}
  {2009})}\BibitemShut {NoStop}%
\bibitem [{\citenamefont {Economou}\ \emph {et~al.}(2010)\citenamefont
  {Economou}, \citenamefont {Lindner},\ and\ \citenamefont {Rudolph}}]{qd2}%
  \BibitemOpen
  \bibfield  {author} {\bibinfo {author} {\bibfnamefont {S.~E.}\ \bibnamefont
  {Economou}}, \bibinfo {author} {\bibfnamefont {N.}~\bibnamefont {Lindner}}, \
  and\ \bibinfo {author} {\bibfnamefont {T.}~\bibnamefont {Rudolph}},\ }\href
  {\doibase 10.1103/PhysRevLett.105.093601} {\bibfield  {journal} {\bibinfo
  {journal} {Phys. Rev. Lett.}\ }\textbf {\bibinfo {volume} {105}},\ \bibinfo
  {pages} {093601} (\bibinfo {year} {2010})}\BibitemShut {NoStop}%
\bibitem [{\citenamefont {Fuchs}\ \emph {et~al.}(2009)\citenamefont {Fuchs},
  \citenamefont {Dobrovitski}, \citenamefont {Toyli}, \citenamefont
  {Heremans},\ and\ \citenamefont {Awschalom}}]{Awschalom_fast_control}%
  \BibitemOpen
  \bibfield  {author} {\bibinfo {author} {\bibfnamefont {G.~D.}\ \bibnamefont
  {Fuchs}}, \bibinfo {author} {\bibfnamefont {V.~V.}\ \bibnamefont
  {Dobrovitski}}, \bibinfo {author} {\bibfnamefont {D.~M.}\ \bibnamefont
  {Toyli}}, \bibinfo {author} {\bibfnamefont {F.~J.}\ \bibnamefont {Heremans}},
  \ and\ \bibinfo {author} {\bibfnamefont {D.~D.}\ \bibnamefont {Awschalom}},\
  }\href {\doibase 10.1126/science.1181193} {\bibfield  {journal} {\bibinfo
  {journal} {Science}\ }\textbf {\bibinfo {volume} {326}},\ \bibinfo {pages}
  {1520} (\bibinfo {year} {2009})}\BibitemShut {NoStop}%
\bibitem [{\citenamefont {Dolde}\ \emph {et~al.}(2014)\citenamefont {Dolde},
  \citenamefont {Bergholm}, \citenamefont {Wang}, \citenamefont {Jakobi},
  \citenamefont {Naydenov}, \citenamefont {Pezzagna}, \citenamefont {Meijer},
  \citenamefont {Jelezko}, \citenamefont {Neumann}, \citenamefont
  {Schulte-Herbruggen}, \citenamefont {Biamonte},\ and\ \citenamefont
  {Wrachtrup}}]{Wrachtrup_opt_control}%
  \BibitemOpen
  \bibfield  {author} {\bibinfo {author} {\bibfnamefont {F.}~\bibnamefont
  {Dolde}}, \bibinfo {author} {\bibfnamefont {V.}~\bibnamefont {Bergholm}},
  \bibinfo {author} {\bibfnamefont {Y.}~\bibnamefont {Wang}}, \bibinfo {author}
  {\bibfnamefont {I.}~\bibnamefont {Jakobi}}, \bibinfo {author} {\bibfnamefont
  {B.}~\bibnamefont {Naydenov}}, \bibinfo {author} {\bibfnamefont
  {S.}~\bibnamefont {Pezzagna}}, \bibinfo {author} {\bibfnamefont
  {J.}~\bibnamefont {Meijer}}, \bibinfo {author} {\bibfnamefont
  {F.}~\bibnamefont {Jelezko}}, \bibinfo {author} {\bibfnamefont
  {P.}~\bibnamefont {Neumann}}, \bibinfo {author} {\bibfnamefont
  {T.}~\bibnamefont {Schulte-Herbruggen}}, \bibinfo {author} {\bibfnamefont
  {J.}~\bibnamefont {Biamonte}}, \ and\ \bibinfo {author} {\bibfnamefont
  {J.}~\bibnamefont {Wrachtrup}},\ }\href {\doibase 10.1038/ncomms4371}
  {\bibfield  {journal} {\bibinfo  {journal} {Nat Commun}\ }\textbf {\bibinfo
  {volume} {5}},\ \bibinfo {pages} {3371} (\bibinfo {year} {2014})}\BibitemShut
  {NoStop}%
\bibitem [{\citenamefont {Neumann}\ \emph {et~al.}(2010)\citenamefont
  {Neumann}, \citenamefont {Beck}, \citenamefont {Steiner}, \citenamefont
  {Rempp}, \citenamefont {Fedder}, \citenamefont {Hemmer}, \citenamefont
  {Wrachtrup},\ and\ \citenamefont {Jelezko}}]{Neumann_singleshot}%
  \BibitemOpen
  \bibfield  {author} {\bibinfo {author} {\bibfnamefont {P.}~\bibnamefont
  {Neumann}}, \bibinfo {author} {\bibfnamefont {J.}~\bibnamefont {Beck}},
  \bibinfo {author} {\bibfnamefont {M.}~\bibnamefont {Steiner}}, \bibinfo
  {author} {\bibfnamefont {F.}~\bibnamefont {Rempp}}, \bibinfo {author}
  {\bibfnamefont {H.}~\bibnamefont {Fedder}}, \bibinfo {author} {\bibfnamefont
  {P.~R.}\ \bibnamefont {Hemmer}}, \bibinfo {author} {\bibfnamefont
  {J.}~\bibnamefont {Wrachtrup}}, \ and\ \bibinfo {author} {\bibfnamefont
  {F.}~\bibnamefont {Jelezko}},\ }\href {\doibase 10.1126/science.1189075}
  {\bibfield  {journal} {\bibinfo  {journal} {Science}\ }\textbf {\bibinfo
  {volume} {329}},\ \bibinfo {pages} {542} (\bibinfo {year}
  {2010})}\BibitemShut {NoStop}%
\bibitem [{\citenamefont {Robledo}\ \emph {et~al.}(2011)\citenamefont
  {Robledo}, \citenamefont {Childress}, \citenamefont {Bernien}, \citenamefont
  {Hensen}, \citenamefont {Alkemade},\ and\ \citenamefont
  {Hanson}}]{Hanson_singleshot}%
  \BibitemOpen
  \bibfield  {author} {\bibinfo {author} {\bibfnamefont {L.}~\bibnamefont
  {Robledo}}, \bibinfo {author} {\bibfnamefont {L.}~\bibnamefont {Childress}},
  \bibinfo {author} {\bibfnamefont {H.}~\bibnamefont {Bernien}}, \bibinfo
  {author} {\bibfnamefont {B.}~\bibnamefont {Hensen}}, \bibinfo {author}
  {\bibfnamefont {P.~F.}\ \bibnamefont {Alkemade}}, \ and\ \bibinfo {author}
  {\bibfnamefont {R.}~\bibnamefont {Hanson}},\ }\href {\doibase
  10.1038/nature10401} {\bibfield  {journal} {\bibinfo  {journal} {Nature}\
  }\textbf {\bibinfo {volume} {477}},\ \bibinfo {pages} {574} (\bibinfo {year}
  {2011})}\BibitemShut {NoStop}%
\bibitem [{\citenamefont {Maurer}\ \emph {et~al.}(2012)\citenamefont {Maurer},
  \citenamefont {Kucsko}, \citenamefont {Latta}, \citenamefont {Jiang},
  \citenamefont {Yao}, \citenamefont {Bennett}, \citenamefont {Pastawski},
  \citenamefont {Hunger}, \citenamefont {Chisholm}, \citenamefont {Markham},
  \citenamefont {Twitchen}, \citenamefont {Cirac},\ and\ \citenamefont
  {Lukin}}]{Lukin_C13}%
  \BibitemOpen
  \bibfield  {author} {\bibinfo {author} {\bibfnamefont {P.~C.}\ \bibnamefont
  {Maurer}}, \bibinfo {author} {\bibfnamefont {G.}~\bibnamefont {Kucsko}},
  \bibinfo {author} {\bibfnamefont {C.}~\bibnamefont {Latta}}, \bibinfo
  {author} {\bibfnamefont {L.}~\bibnamefont {Jiang}}, \bibinfo {author}
  {\bibfnamefont {N.~Y.}\ \bibnamefont {Yao}}, \bibinfo {author} {\bibfnamefont
  {S.~D.}\ \bibnamefont {Bennett}}, \bibinfo {author} {\bibfnamefont
  {F.}~\bibnamefont {Pastawski}}, \bibinfo {author} {\bibfnamefont
  {D.}~\bibnamefont {Hunger}}, \bibinfo {author} {\bibfnamefont
  {N.}~\bibnamefont {Chisholm}}, \bibinfo {author} {\bibfnamefont
  {M.}~\bibnamefont {Markham}}, \bibinfo {author} {\bibfnamefont {D.~J.}\
  \bibnamefont {Twitchen}}, \bibinfo {author} {\bibfnamefont {J.~I.}\
  \bibnamefont {Cirac}}, \ and\ \bibinfo {author} {\bibfnamefont {M.~D.}\
  \bibnamefont {Lukin}},\ }\href {\doibase 10.1126/science.1220513} {\bibfield
  {journal} {\bibinfo  {journal} {Science}\ }\textbf {\bibinfo {volume}
  {336}},\ \bibinfo {pages} {1283} (\bibinfo {year} {2012})}\BibitemShut
  {NoStop}%
\bibitem [{\citenamefont {Waldherr}\ \emph {et~al.}(2014)\citenamefont
  {Waldherr}, \citenamefont {Wang}, \citenamefont {Zaiser}, \citenamefont
  {Jamali}, \citenamefont {Schulte-Herbruggen}, \citenamefont {Abe},
  \citenamefont {Ohshima}, \citenamefont {Isoya}, \citenamefont {Du},
  \citenamefont {Neumann},\ and\ \citenamefont
  {Wrachtrup}}]{Wrachtrup_error_correction}%
  \BibitemOpen
  \bibfield  {author} {\bibinfo {author} {\bibfnamefont {G.}~\bibnamefont
  {Waldherr}}, \bibinfo {author} {\bibfnamefont {Y.}~\bibnamefont {Wang}},
  \bibinfo {author} {\bibfnamefont {S.}~\bibnamefont {Zaiser}}, \bibinfo
  {author} {\bibfnamefont {M.}~\bibnamefont {Jamali}}, \bibinfo {author}
  {\bibfnamefont {T.}~\bibnamefont {Schulte-Herbruggen}}, \bibinfo {author}
  {\bibfnamefont {H.}~\bibnamefont {Abe}}, \bibinfo {author} {\bibfnamefont
  {T.}~\bibnamefont {Ohshima}}, \bibinfo {author} {\bibfnamefont
  {J.}~\bibnamefont {Isoya}}, \bibinfo {author} {\bibfnamefont {J.~F.}\
  \bibnamefont {Du}}, \bibinfo {author} {\bibfnamefont {P.}~\bibnamefont
  {Neumann}}, \ and\ \bibinfo {author} {\bibfnamefont {J.}~\bibnamefont
  {Wrachtrup}},\ }\href {\doibase 10.1038/nature12919} {\bibfield  {journal}
  {\bibinfo  {journal} {Nature}\ }\textbf {\bibinfo {volume} {506}},\ \bibinfo
  {pages} {204} (\bibinfo {year} {2014})}\BibitemShut {NoStop}%
\bibitem [{\citenamefont {Dolde}\ \emph {et~al.}(2013)\citenamefont {Dolde},
  \citenamefont {Jakobi}, \citenamefont {Naydenov}, \citenamefont {Zhao},
  \citenamefont {Pezzagna}, \citenamefont {Trautmann}, \citenamefont {Meijer},
  \citenamefont {Neumann}, \citenamefont {Jelezko},\ and\ \citenamefont
  {Wrachtrup}}]{Wrachtrup_pairs}%
  \BibitemOpen
  \bibfield  {author} {\bibinfo {author} {\bibfnamefont {F.}~\bibnamefont
  {Dolde}}, \bibinfo {author} {\bibfnamefont {I.}~\bibnamefont {Jakobi}},
  \bibinfo {author} {\bibfnamefont {B.}~\bibnamefont {Naydenov}}, \bibinfo
  {author} {\bibfnamefont {N.}~\bibnamefont {Zhao}}, \bibinfo {author}
  {\bibfnamefont {S.}~\bibnamefont {Pezzagna}}, \bibinfo {author}
  {\bibfnamefont {C.}~\bibnamefont {Trautmann}}, \bibinfo {author}
  {\bibfnamefont {J.}~\bibnamefont {Meijer}}, \bibinfo {author} {\bibfnamefont
  {P.}~\bibnamefont {Neumann}}, \bibinfo {author} {\bibfnamefont
  {F.}~\bibnamefont {Jelezko}}, \ and\ \bibinfo {author} {\bibfnamefont
  {J.}~\bibnamefont {Wrachtrup}},\ }\href {\doibase
  http://www.nature.com/nphys/journal/v9/n3/abs/nphys2545.html#supplementary-information}
  {\bibfield  {journal} {\bibinfo  {journal} {Nat Phys}\ }\textbf {\bibinfo
  {volume} {9}},\ \bibinfo {pages} {139} (\bibinfo {year} {2013})}\BibitemShut
  {NoStop}%
\bibitem [{\citenamefont {Taminiau}\ \emph {et~al.}(2014)\citenamefont
  {Taminiau}, \citenamefont {Cramer}, \citenamefont {van~der Sar},
  \citenamefont {Dobrovitski},\ and\ \citenamefont
  {Hanson}}]{Hanson_error_correction}%
  \BibitemOpen
  \bibfield  {author} {\bibinfo {author} {\bibfnamefont {T.~H.}\ \bibnamefont
  {Taminiau}}, \bibinfo {author} {\bibfnamefont {J.}~\bibnamefont {Cramer}},
  \bibinfo {author} {\bibfnamefont {T.}~\bibnamefont {van~der Sar}}, \bibinfo
  {author} {\bibfnamefont {V.~V.}\ \bibnamefont {Dobrovitski}}, \ and\ \bibinfo
  {author} {\bibfnamefont {R.}~\bibnamefont {Hanson}},\ }\href {\doibase
  10.1038/nnano.2014.2} {\bibfield  {journal} {\bibinfo  {journal} {Nat
  Nanotechnol}\ }\textbf {\bibinfo {volume} {9}},\ \bibinfo {pages} {171}
  (\bibinfo {year} {2014})}\BibitemShut {NoStop}%
\bibitem [{\citenamefont {Togan}\ \emph {et~al.}(2010)\citenamefont {Togan},
  \citenamefont {Chu}, \citenamefont {Trifonov}, \citenamefont {Jiang},
  \citenamefont {Maze}, \citenamefont {Childress}, \citenamefont {Dutt},
  \citenamefont {Sorensen}, \citenamefont {Hemmer}, \citenamefont {Zibrov},\
  and\ \citenamefont {Lukin}}]{Lukin_photon}%
  \BibitemOpen
  \bibfield  {author} {\bibinfo {author} {\bibfnamefont {E.}~\bibnamefont
  {Togan}}, \bibinfo {author} {\bibfnamefont {Y.}~\bibnamefont {Chu}}, \bibinfo
  {author} {\bibfnamefont {A.~S.}\ \bibnamefont {Trifonov}}, \bibinfo {author}
  {\bibfnamefont {L.}~\bibnamefont {Jiang}}, \bibinfo {author} {\bibfnamefont
  {J.}~\bibnamefont {Maze}}, \bibinfo {author} {\bibfnamefont {L.}~\bibnamefont
  {Childress}}, \bibinfo {author} {\bibfnamefont {M.~V.}\ \bibnamefont {Dutt}},
  \bibinfo {author} {\bibfnamefont {A.~S.}\ \bibnamefont {Sorensen}}, \bibinfo
  {author} {\bibfnamefont {P.~R.}\ \bibnamefont {Hemmer}}, \bibinfo {author}
  {\bibfnamefont {A.~S.}\ \bibnamefont {Zibrov}}, \ and\ \bibinfo {author}
  {\bibfnamefont {M.~D.}\ \bibnamefont {Lukin}},\ }\href {\doibase
  10.1038/nature09256} {\bibfield  {journal} {\bibinfo  {journal} {Nature}\
  }\textbf {\bibinfo {volume} {466}},\ \bibinfo {pages} {730} (\bibinfo {year}
  {2010})}\BibitemShut {NoStop}%
\bibitem [{\citenamefont {Kosaka}\ and\ \citenamefont
  {Niikura}(2015)}]{kosaka_entangled}%
  \BibitemOpen
  \bibfield  {author} {\bibinfo {author} {\bibfnamefont {H.}~\bibnamefont
  {Kosaka}}\ and\ \bibinfo {author} {\bibfnamefont {N.}~\bibnamefont
  {Niikura}},\ }\href {\doibase 10.1103/PhysRevLett.114.053603} {\bibfield
  {journal} {\bibinfo  {journal} {Phys. Rev. Lett.}\ }\textbf {\bibinfo
  {volume} {114}},\ \bibinfo {pages} {053603} (\bibinfo {year}
  {2015})}\BibitemShut {NoStop}%
\bibitem [{\citenamefont {Bernien}\ \emph {et~al.}(2013)\citenamefont
  {Bernien}, \citenamefont {Hensen}, \citenamefont {Pfaff}, \citenamefont
  {Koolstra}, \citenamefont {Blok}, \citenamefont {Robledo}, \citenamefont
  {Taminiau}, \citenamefont {Markham}, \citenamefont {Twitchen}, \citenamefont
  {Childress},\ and\ \citenamefont {Hanson}}]{Hanson_entanglement}%
  \BibitemOpen
  \bibfield  {author} {\bibinfo {author} {\bibfnamefont {H.}~\bibnamefont
  {Bernien}}, \bibinfo {author} {\bibfnamefont {B.}~\bibnamefont {Hensen}},
  \bibinfo {author} {\bibfnamefont {W.}~\bibnamefont {Pfaff}}, \bibinfo
  {author} {\bibfnamefont {G.}~\bibnamefont {Koolstra}}, \bibinfo {author}
  {\bibfnamefont {M.~S.}\ \bibnamefont {Blok}}, \bibinfo {author}
  {\bibfnamefont {L.}~\bibnamefont {Robledo}}, \bibinfo {author} {\bibfnamefont
  {T.~H.}\ \bibnamefont {Taminiau}}, \bibinfo {author} {\bibfnamefont
  {M.}~\bibnamefont {Markham}}, \bibinfo {author} {\bibfnamefont {D.~J.}\
  \bibnamefont {Twitchen}}, \bibinfo {author} {\bibfnamefont {L.}~\bibnamefont
  {Childress}}, \ and\ \bibinfo {author} {\bibfnamefont {R.}~\bibnamefont
  {Hanson}},\ }\href {\doibase 10.1038/nature12016
  http://www.nature.com/nature/journal/v497/n7447/abs/nature12016.html#supplementary-information}
  {\bibfield  {journal} {\bibinfo  {journal} {Nature}\ }\textbf {\bibinfo
  {volume} {497}},\ \bibinfo {pages} {86} (\bibinfo {year} {2013})}\BibitemShut
  {NoStop}%
\bibitem [{\citenamefont {Pfaff}\ \emph {et~al.}(2014)\citenamefont {Pfaff},
  \citenamefont {Hensen}, \citenamefont {Bernien}, \citenamefont {van Dam},
  \citenamefont {Blok}, \citenamefont {Taminiau}, \citenamefont {Tiggelman},
  \citenamefont {Schouten}, \citenamefont {Markham}, \citenamefont {Twitchen},\
  and\ \citenamefont {Hanson}}]{Hanson_teleportation}%
  \BibitemOpen
  \bibfield  {author} {\bibinfo {author} {\bibfnamefont {W.}~\bibnamefont
  {Pfaff}}, \bibinfo {author} {\bibfnamefont {B.~J.}\ \bibnamefont {Hensen}},
  \bibinfo {author} {\bibfnamefont {H.}~\bibnamefont {Bernien}}, \bibinfo
  {author} {\bibfnamefont {S.~B.}\ \bibnamefont {van Dam}}, \bibinfo {author}
  {\bibfnamefont {M.~S.}\ \bibnamefont {Blok}}, \bibinfo {author}
  {\bibfnamefont {T.~H.}\ \bibnamefont {Taminiau}}, \bibinfo {author}
  {\bibfnamefont {M.~J.}\ \bibnamefont {Tiggelman}}, \bibinfo {author}
  {\bibfnamefont {R.~N.}\ \bibnamefont {Schouten}}, \bibinfo {author}
  {\bibfnamefont {M.}~\bibnamefont {Markham}}, \bibinfo {author} {\bibfnamefont
  {D.~J.}\ \bibnamefont {Twitchen}}, \ and\ \bibinfo {author} {\bibfnamefont
  {R.}~\bibnamefont {Hanson}},\ }\href {\doibase 10.1126/science.1253512}
  {\bibfield  {journal} {\bibinfo  {journal} {Science}\ }\textbf {\bibinfo
  {volume} {345}},\ \bibinfo {pages} {532} (\bibinfo {year}
  {2014})}\BibitemShut {NoStop}%
\bibitem [{\citenamefont {Nemoto}\ \emph {et~al.}(2014)\citenamefont {Nemoto},
  \citenamefont {Trupke}, \citenamefont {Devitt}, \citenamefont {Stephens},
  \citenamefont {Scharfenberger}, \citenamefont {Buczak}, \citenamefont
  {N\"obauer}, \citenamefont {Everitt}, \citenamefont {Schmiedmayer},\ and\
  \citenamefont {Munro}}]{nuclearcluster}%
  \BibitemOpen
  \bibfield  {author} {\bibinfo {author} {\bibfnamefont {K.}~\bibnamefont
  {Nemoto}}, \bibinfo {author} {\bibfnamefont {M.}~\bibnamefont {Trupke}},
  \bibinfo {author} {\bibfnamefont {S.~J.}\ \bibnamefont {Devitt}}, \bibinfo
  {author} {\bibfnamefont {A.~M.}\ \bibnamefont {Stephens}}, \bibinfo {author}
  {\bibfnamefont {B.}~\bibnamefont {Scharfenberger}}, \bibinfo {author}
  {\bibfnamefont {K.}~\bibnamefont {Buczak}}, \bibinfo {author} {\bibfnamefont
  {T.}~\bibnamefont {N\"obauer}}, \bibinfo {author} {\bibfnamefont {M.~S.}\
  \bibnamefont {Everitt}}, \bibinfo {author} {\bibfnamefont {J.}~\bibnamefont
  {Schmiedmayer}}, \ and\ \bibinfo {author} {\bibfnamefont {W.~J.}\
  \bibnamefont {Munro}},\ }\href {\doibase 10.1103/PhysRevX.4.031022}
  {\bibfield  {journal} {\bibinfo  {journal} {Phys. Rev. X}\ }\textbf {\bibinfo
  {volume} {4}},\ \bibinfo {pages} {031022} (\bibinfo {year}
  {2014})}\BibitemShut {NoStop}%
\bibitem [{\citenamefont {Maze}\ \emph {et~al.}(2011)\citenamefont {Maze},
  \citenamefont {Gali}, \citenamefont {Togan}, \citenamefont {Chu},
  \citenamefont {Trifonov}, \citenamefont {Kaxiras},\ and\ \citenamefont
  {Lukin}}]{Maze_NV_structure}%
  \BibitemOpen
  \bibfield  {author} {\bibinfo {author} {\bibfnamefont {J.~R.}\ \bibnamefont
  {Maze}}, \bibinfo {author} {\bibfnamefont {A.}~\bibnamefont {Gali}}, \bibinfo
  {author} {\bibfnamefont {E.}~\bibnamefont {Togan}}, \bibinfo {author}
  {\bibfnamefont {Y.}~\bibnamefont {Chu}}, \bibinfo {author} {\bibfnamefont
  {A.}~\bibnamefont {Trifonov}}, \bibinfo {author} {\bibfnamefont
  {E.}~\bibnamefont {Kaxiras}}, \ and\ \bibinfo {author} {\bibfnamefont
  {M.~D.}\ \bibnamefont {Lukin}},\ }\href {\doibase
  10.1088/1367-2630/13/2/025025} {\bibfield  {journal} {\bibinfo  {journal}
  {New Journal of Physics}\ }\textbf {\bibinfo {volume} {13}},\ \bibinfo
  {pages} {025025} (\bibinfo {year} {2011})}\BibitemShut {NoStop}%
\bibitem [{\citenamefont {Briegel}\ and\ \citenamefont
  {Raussendorf}(2001)}]{cluster}%
  \BibitemOpen
  \bibfield  {author} {\bibinfo {author} {\bibfnamefont {H.~J.}\ \bibnamefont
  {Briegel}}\ and\ \bibinfo {author} {\bibfnamefont {R.}~\bibnamefont
  {Raussendorf}},\ }\href {\doibase 10.1103/PhysRevLett.86.910} {\bibfield
  {journal} {\bibinfo  {journal} {Phys. Rev. Lett.}\ }\textbf {\bibinfo
  {volume} {86}},\ \bibinfo {pages} {910} (\bibinfo {year} {2001})}\BibitemShut
  {NoStop}%
\bibitem [{\citenamefont {Faraon}\ \emph {et~al.}(2012)\citenamefont {Faraon},
  \citenamefont {Santori}, \citenamefont {Huang}, \citenamefont {Acosta},\ and\
  \citenamefont {Beausoleil}}]{cavity}%
  \BibitemOpen
  \bibfield  {author} {\bibinfo {author} {\bibfnamefont {A.}~\bibnamefont
  {Faraon}}, \bibinfo {author} {\bibfnamefont {C.}~\bibnamefont {Santori}},
  \bibinfo {author} {\bibfnamefont {Z.}~\bibnamefont {Huang}}, \bibinfo
  {author} {\bibfnamefont {V.~M.}\ \bibnamefont {Acosta}}, \ and\ \bibinfo
  {author} {\bibfnamefont {R.~G.}\ \bibnamefont {Beausoleil}},\ }\href
  {\doibase 10.1103/PhysRevLett.109.033604} {\bibfield  {journal} {\bibinfo
  {journal} {Phys. Rev. Lett.}\ }\textbf {\bibinfo {volume} {109}},\ \bibinfo
  {pages} {033604} (\bibinfo {year} {2012})}\BibitemShut {NoStop}%
\bibitem [{\citenamefont {Arroyo-Camejo}\ \emph {et~al.}(2014)\citenamefont
  {Arroyo-Camejo}, \citenamefont {Lazariev}, \citenamefont {Hell},\ and\
  \citenamefont {Balasubramanian}}]{holonomic}%
  \BibitemOpen
  \bibfield  {author} {\bibinfo {author} {\bibfnamefont {S.}~\bibnamefont
  {Arroyo-Camejo}}, \bibinfo {author} {\bibfnamefont {A.}~\bibnamefont
  {Lazariev}}, \bibinfo {author} {\bibfnamefont {S.~W.}\ \bibnamefont {Hell}},
  \ and\ \bibinfo {author} {\bibfnamefont {G.}~\bibnamefont
  {Balasubramanian}},\ }\href@noop {} {\bibfield  {journal} {\bibinfo
  {journal} {Nat Commun}\ }\textbf {\bibinfo {volume} {5}},\ \bibinfo {pages}
  {4870} (\bibinfo {year} {2014})}\BibitemShut {NoStop}%
\end{thebibliography}%

\end{document}